# A macro agent and its actions

Larissa Albantakis[1,*], Francesco Massari[2,&], Maggie Beheler-Amass[1,&], and Giulio Tononi[1]

[1] Department of Psychiatry, Wisconsin Institute for Sleep and Consciousness, University of Wisconsin-Madison, Madison, WI 53715, USA
[2] Swarthmore College, Swarthmore, PA 19081, USA
* Correspondence: albantakis@wisc.edu
[&] These authors contributed equally to this work.

## Abstract

In science, macro level descriptions of the causal interactions within complex, dynamical systems are typically deemed convenient, but ultimately reducible to a complete causal account of the underlying micro constituents. Yet, such a reductionist perspective is hard to square with several issues related to autonomy and agency: (1) agents require (causal) borders that separate them from the environment, (2) at least in a biological context, agents are associated with macroscopic systems, and (3) agents are supposed to act upon their environment. Integrated information theory (IIT) (Oizumi et al., 2014) offers a quantitative account of causation based on a set of causal principles, including notions such as causal specificity, composition, and irreducibility, that challenges the reductionist perspective in multiple ways. First, the IIT formalism provides a complete account of a system's causal structure, including irreducible higher-order mechanisms constituted of multiple system elements. Second, a system's amount of integrated information ($\Phi$) measures the causal constraints a system exerts onto itself and can peak at a macro level of description (Hoel et al., 2016; Marshall et al., 2018). Finally, the causal principles of IIT can also be employed to identify and quantify the actual causes of events ("what caused what"), such as an agent's actions (Albantakis et al., 2019). Here, we demonstrate this framework by example of a simulated agent, equipped with a small neural network, that forms a maximum of $\Phi$ at a macro scale.

## Introduction

What is an agent? To date, there is no single, agreed-upon definition of an agent that captures all relevant intuitions behind the concept. As a minimal property, an agent must be an open system that dynamically and informationally interacts with an environment. This simple requirement, however, immediately poses a methodological problem: When subsystems within a larger system are characterized by biological or informational properties, their boundaries are typically taken for granted and assumed as given (Krakauer et al., 2014; Oizumi et al., 2014; Albantakis, 2018; Kolchinsky and Wolpert, 2018).

Moreover, at least in a biological context, the notion of agency is typically associated with macroscopic spatio-temporal scales. For example, the causally relevant information that organisms pick up from their environment is generally not dependent on microscopic details; the cognitive abilities of an animal are related to the interactions of their neurons, rather than the underlying molecules, atoms or quarks; and the goal-directed actions performed by humans and other animals, and even life itself, have been characterized as instances of top-down causation (Ellis, 2009, 2016; Walker and Davies, 2013).

This brings us to the last point: that agents are supposed to act upon their environment. Yet, as many have argued, the fact that physical events are either determined by previous (micro-physical) events, or emerge from (quantum) randomness, seems to be at odds with the notion of an autonomous agent with intrinsic causal power.

Originally developed as a theory of consciousness (Tononi, 2015; Tononi et al., 2016), IIT offers a quantitative framework to characterize the causal structure of discrete dynamical systems (Oizumi et al., 2014). The main quantity, $\Phi$, measures to what extent the causal constraints that a system exerts onto itself are irreducible to those of its parts. A system with $\Phi > 0$, forms a unitary whole, as all of its subsets constrain and are being constrained by other subsets within the system above a background of external influences (Maturana and Varela, 1980; Tononi, 2013; Marshall et al., 2017; Aguilera and Di Paolo, 2018; Albantakis, 2018; Farnsworth, 2018). In this way, IIT provides the tools to identify whether a set of elements forms an entity with causal borders that separate it from its environment – a maximum of $\Phi$ (Oizumi et al., 2014; Marshall et al., 2017).

The IIT formalism can also be applied across micro and macro spatiotemporal scales in order to identify those organizational levels at which the system exhibits strong causal constraints onto itself. As shown in previous work (Hoel et al., 2016; Marshall et al., 2018), it is indeed possible for a system to have higher $\Phi$ at a macro level than at the micro level. According to IIT principles, the particular spatio-temporal scale that specifies a maximum of integrated information ($\Phi^{max}$) defines the spatio-temporal scale at which the system specifies itself in causal terms.

Finally, the causal principles of IIT, including notions such as causal specificity, composition, and irreducibility, can be employed to identify and quantify the actual causes and effects of events ("what caused what") within a transition between subsequent states of discrete dynamical system (Albantakis et al., 2019). Such a principled account of actual causation makes it possible to identify the causes of an agent's actions and to trace them back in time ("causes of causes") (Juel et al., 2019).

Our goal here is to demonstrate how various aspects of the IIT formalism can be combined to provide an account of (macro) agents and their actions *in silico*, through the example of a simulated agent ("animat") equipped with a small neural network that is able to perform a simple perceptual categorization task (Figure 1) (Beer, 2003; Albantakis et al., 2014; Hintze et al., 2017). We will first describe the animat as a macro system of interacting "neurons" (black boxes) (Marshall et al., 2018), before zooming in on its micro constituents (Figure 2). As we will show, the animat exhibits higher values of integrated information ($\Phi$) at the macro level than at the micro level. Next, we will trace the causes of the animat's actions back in time. While each action is necessarily preceded by a chain of micro events, these can only account for parts of the action. In our example, a cause for the action as a higher-order event constituted of multiple micro occurrences can only be found at the macro spatio-temporal scale. More broadly, our example analysis serves to demonstrate that IIT's causal framework provides a consistent, quantitative account of causation that challenges the wide-spread reductionist perspective—that only individual micro constituents ultimately have causal power.

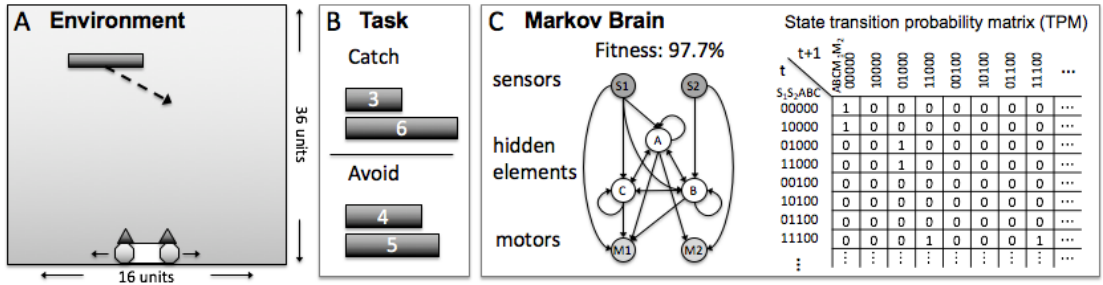

*Figure 1. Artificial agent ("animat") capable of performing a perceptual categorization task. (A) The animat is placed in a 16 by 36 unit environment. The animat itself is 3 units wide and can move to the right and left 1 unit at a time. Its two sensors are positioned on each side with 1 unit between them and switch 'on' (1) whenever a block is directly above them irrespective of its distance. Per trial one block is falling to the right or left at one unit per time step. (B) The task is to catch blocks of size 3 and 6 and to avoid blocks of size 4 and 5 (Task 4 in (Albantakis et al., 2014)). (C) The animat is equipped with a binary, deterministic Markov Brain (Hintze et al., 2017) constituted—at the macro level—of two sensors, 3 hidden elements, and two motors. The update function is fully described by the animat's deterministic state transition probability matrix. The sensor states are determined by the environment. The motor elements do not have feedback connections to the rest of the system.*

## The simulated animat – macro and micro

The animat we will analyze in the following is capable of performing an active perceptual categorization task (Beer, 2003; Marstaller et al., 2013) with high accuracy (97.7% correct). In the simulated environment, blocks of different sizes are falling to the right or left, one at a time, and the animat has to catch or avoid them depending on their size (Figure 1). To that end, the animat is equipped with two sensors that turn 'on' (1) if a block is positioned directly above them at any distance, and two motors that enable the animat to move to the right or left ($M_1 M_2 = (0,1)$: move right, $M_1 M_2 = (1,0)$: move left, $M_1 M_2 = \{(0,0), (1,1)\}$: stand still). The animat's behavior is determined by a "Markov Brain" (Hintze et al., 2017), a small neural network which, in our specific case, is constituted of binary elements with deterministic input-output functions. In (Albantakis 2014), we have used a genetic algorithm to evolve a population of this type of Markov Brains for high fitness in the block-catching task. Both the connectivity structure and update function of the Markov Brains were encoded in a genome (string of integers) and adapted through fitness selection and mutation.

***Macro level network:*** At the macro level (Figure 1C), the Markov Brain of our example animat is equivalent to that of the best performing animat in (Albantakis 2014) (Task 4), with three hidden nodes $(A, B, C)$ that are connected in an all-to-all manner, while connections from the sensors and to the motors are feedforward only. The update function of the animat's Markov Brain can be represented by its state transition probability matrix (TPM), which specifies the output state of the hidden nodes and motors given the prior state of the sensors and the hidden nodes (the state of the sensors is fully determined by the environment). The macro elements, nodes $A, B, C$ and motors $M_1, M_2$, update at the same rate as the environment.

By construction, each macro element of our example animat corresponds to a "black box" (Marshall et al., 2018), constituted of a set of micro elements that interact and update at a finer spatial and temporal scale. In Figure 2 we zoom in on the micro constituents of the macro-level animat displayed in Figure 1C. Within the IIT framework, the macro level strictly supervenes upon its micro constituents. The macro level corresponds to a mapping that groups disjoint subsets of micro elements into non-overlapping macro elements (Hoel et al., 2013, 2016; Marshall et al., 2018). Likewise, the state of a macro element is always determined by a surjective mapping of the micro states of its underlying micro constituents.

*Micro level network:* The animat's micro level is constituted of 72 simple logic gates (COPY, AND, OR, XOR, NOT, and NOR gates, as well as one Majority (MAJ) gate that turns 'on' if more than half of its inputs are 'on'). As shown in Figure 2, the sensors consist of the same two elements at the macro and micro level. However, the macro elements $A, B, C, M_1, M_2$ each correspond to a "black box" (Marshall et al., 2018) constituted of several (10 to 17) micro elements. The micro elements within these black boxes are connected in a largely feed-forward manner (except for one loop in each motor black box, which "clocks" its motor response (see below)) and collectively emulate the logic function of their respective macro node over 4 micro time steps (updates). Each black box has one output node, but may receive inputs from the other black boxes or the sensors via multiple input nodes. The output nodes of the motor black boxes determine the animat's movements but do not feedback into the network. In this way, the connectivity between the black boxes mirrors the connectivity of the macro-level animat shown in Figure 1C.

*State-mapping:* As each black-box has four layers (including inputs and the output node), four micro updates correspond to one macro update. The environment updates at the same rate as the macro elements. This means that, when viewed at the micro level, the sensors receive and output the same environmental input for four micro time steps.

The mapping from micro-level states to macro-level states is accomplished as proposed by (Marshall et al., 2018). The macro state of a black box corresponds to the state of its output node at the time of the macro update (here, every four micro updates) (Figure 2, bottom). The states of all non-output elements are ignored at the macro level, as they are hidden within the black boxes. As a consequence, each macro state is realizable by multiple micro states. Figure 2 shows one possible micro state corresponding to the macro state $S_1 S_2 A B C M_1 M_2 = (0,0,1,0,1,1,0)$. Similarly, the states between the macro updates are ignored for determining the macro update function.

Due to the specific implementation of the motor black boxes, their outputs remain in state $(0,0)$ between macro updates, if the Markov Brain is initialized correctly at the beginning of each trial (e.g., in state "all off"). This means that the animat may only perform actions on those micro time steps that correspond to the macro update. This guarantees that our example animat behaves in exactly the same way as an animat that, at the micro level, is implemented as in Figure 1C without further sub-constituents.

In the following we will apply IIT's causal analysis to our example animat, at both the macro and the micro level. To that end, we will first evaluate the causal constraints the system exerts onto itself—its *cause-effect structure*—at both levels. Second, we will assess to what extent the constraints specified by the cause-effect structure are irreducible under a partition of the system, as measured by $\Phi$.

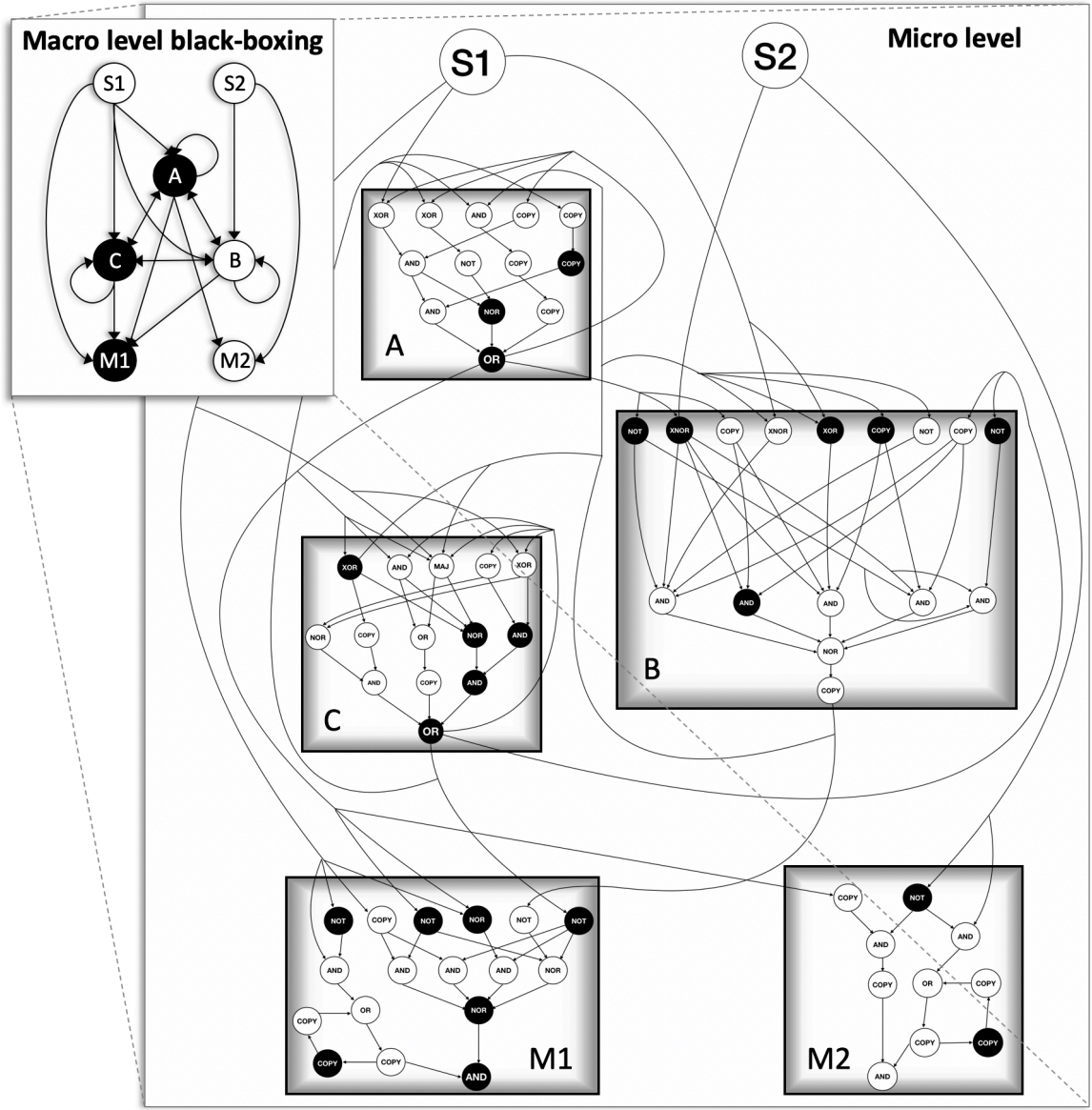

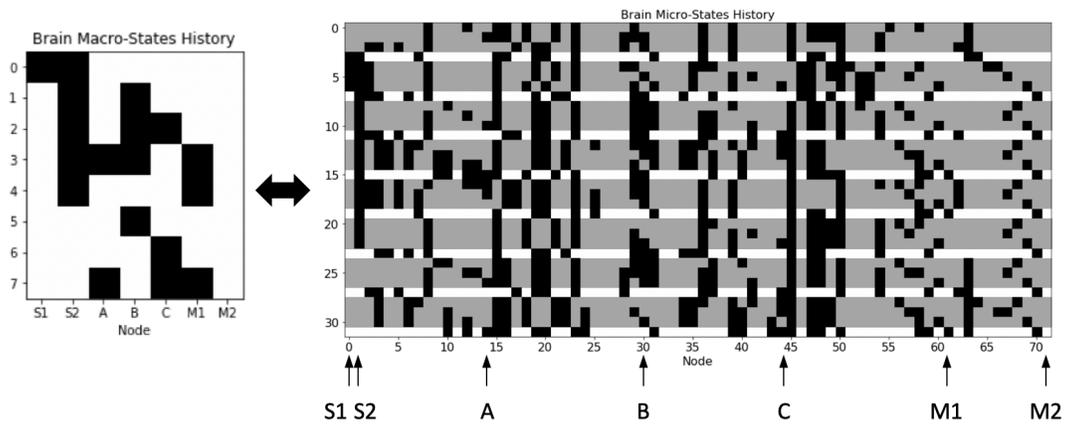

*Figure 2. Looking inside the macro level black-boxing at the animat's micro level constituents.* Each macro element $A, B, C, M_1, M_2$ is constituted of several (10 to 17) micro elements (simple logic gates: COPY, AND, OR, XOR, NOT, NOR, and MAJ as indicated). Within each black box, micro elements are connected in a largely feed-forward manner (except for one loop in each motor black box). Every four micro updates correspond to one macro update. The macro state, here $S_1 S_2 A B C M_1 M_2 = (0,0,1,0,1,1,0)$, corresponds to the state of the black-box output nodes at the time of each macro update and is thus multiply realizable. Shown here is one possible realization corresponding to the last state in the time series shown below. The states of the other micro elements are ignored at the macro level (as they are hidden within the black boxes). Likewise, the state of the other micro timesteps are not taken into account in the mapping (Marshall et al., 2018). Given the particular implementation of the motor black boxes, the animat may only move on those micro time steps that correspond to the macro updates.

## The compositional cause-effect structure of a system in a state

The IIT formalism evaluates five causal principles: intrinsicality, composition, information, integration, and exclusion (Oizumi et al., 2014; Tononi, 2015). We will briefly outline these principles underlying IIT's causal analysis by example of the macro animat shown in Figure 1C and its corresponding transition probability matrix (TPM). For details and formal definitions of the relevant quantities we refer to the original publications (Oizumi et al., 2014; Tononi, 2015). All IIT quantities can be computed from a given TPM using PyPhi, IIT's python toolbox (Mayner et al., 2018). Here, we used the standard configuration corresponding to "IIT 3.0" (Oizumi et al., 2014).

In general, the IIT analysis starts from a discrete dynamical system $S$, constituted of $n$ interacting elements $S_i$ with $i = 1, ..., n$. Each element must have at least two internal states, which can be observed and manipulated, and is equipped with a Markovian input-output function $f_i$ that determines the element's output state $s_{i,t}$ depending only on the previous system state $s_{t-1}$: $s_{i,t} = f_i(S_{t-1} = s_{t-1})$. This means that all elements are conditionally independent given the past state $s_{t-1}$ of the system. $S$ is fully described by its state transition probabilities:

$$\hat{p}(S_t = s_t | S_{t-1} = s_{t-1}) = \prod_{i=1}^{n} \hat{p}(S_{i,t} = s_{i,t} | S_{t-1} = s_{t-1}), \quad \forall s_t, s_{t-1}. \quad (1)$$

Note that Eqn. 1 includes system states that may not be observed during the dynamical evolution of the system, but require system interventions (Pearl, 2000; Ay and Polani, 2008). The notation $\hat{p}$ emphasizes that all probabilities herein correspond to interventions, not mere observations: $\hat{p}(S_t = s_t | S_{t-1} = s_{t-1}) = p(S_t = s_t | do(S_{t-1} = s_{t-1}))$ (Pearl, 2000).

*Intrinsicality:* Our goal is to evaluate the causal constraints specified by the set of elements $S$ onto itself, above the background of external influences. If $S$ is a subset of elements within a larger system, all elements outside of $S$ are held fixed in their current state throughout the causal analysis and thus act as background conditions (causal conditioning). From its intrinsic perspective, the system is always in one particular state at any given moment. Accordingly, IIT's causal analysis is state-dependent—it characterizes the system in its current state—and we take all previous system states $s_{t-1}$ to be *a priori* equally probable (maximum entropy). State-averaged system properties, such as its stationary distribution, or an observed time-series, are extrinsic, available to an external observer but not the system itself.

For illustration, we choose the macro candidate set $ABC$ in state $ABC = (1,0,1)$ (Figure 3A). In that case, the sensors and motors $S_1 S_2 M_1 M_2 = (0,0,1,0)$ act as fixed background conditions. The transition probabilities of $ABC$ can be obtained from the full TPM (Figure 1C) by conditioning on $S_1 S_2 M_1 M_2 = (0,0,1,0)$ and are shown in Figure 3B. While the macro level TPM supervenes upon the micro level TPM, we ignore the underlying micro updates here and only take the macro TPM into account when we assess the animat's macro cause-effect structure. From the intrinsic perspective of the system at the macro level, the micro elements and their updates are hidden inside the black boxes. Only the constraints between the macro elements should be taken into account.

***Composition:*** In contrast with reductionist accounts that only consider how individual system elements update and interact, and holistic approaches that describe the dynamical evolution of the system as a whole based on its global state transitions, IIT takes a compositional perspective on causation (Albantakis and Tononi, 2019). Not only single elements (here $A$, $B$, and $C$), but also combinations of elements may specify their own constraints about other system subsets as long as they are irreducible (see below). Within our candidate set $ABC = (1,0,1)$ we thus evaluate the integrated information $\varphi(x_t)$ of all subsets $X = x_t$ of $ABC = (1,0,1)$ (Figure 3C). A subset $X$ with $\varphi(x_t) > 0$ is termed a mechanism within the system in its current state $S = s_t$. Mechanisms constituted of single elements are termed "first-order mechanisms", while those constituted of multiple elements are termed "higher-order mechanisms" and are occasionally labeled by their specific order, e.g., "second-order mechanism" for $AC = (1,1)$.

***Information:*** The IIT formalism employs a counterfactual, interventionist notion of causation (Lewis, 1973; Pearl, 2000) to evaluate the causal constraints that a set of elements in its current state specifies about its causes and effects within the system. However, rather than testing for a counterfactual relation based on a single alternative, IIT considers all possible system states in its causal analysis, which can thus be expressed in probabilistic, informational terms (Albantakis et al., 2019). For clarity we add a time subscript to indicate a system subset at a specific point in time. The constraints that a system subset $X_t \subseteq S$ in its current state $x_t \subseteq s_t$ specifies about the prior or next state of another subset $Z_{t\pm1} \subseteq S$ are captured by its cause or effect repertoire. Specifically, the effect repertoire of $x_t$ over the subset $Z_{t+1}$ is defined as:

$$\pi(Z_{t+1}|x_t) = \prod_i \hat{p}(Z_{i,t+1}|x_t). \qquad (2)$$

The symbol $\pi$ indicates that the repertoire is a product distribution over the individual elements $Z_{i,t+1} \in Z_{t+1}$ rather than simply the conditional distribution over $Z_{t+1}$. In this way, all $Z_{i,t+1}$ are conditioned on $x_t$ but receive independent "random" inputs from variables in $S_t \setminus X_t$ which are marginalized (causal marginalization). The cause repertoire of $x_t$ over the subset $Z_{t-1}$ is defined as:

$$\pi(Z_{t-1}|x_t) = \frac{1}{K}\prod_i \hat{p}(Z_{t-1}|x_{i,t}) \quad \text{with} \quad K = \sum_{z \in Z}\prod_i \hat{p}(Z_{t-1} = z|x_{i,t}). \qquad (3)$$

Here the product is over the elements in $x_t$, which discounts biases from common inputs from $S_{t-1} \setminus Z_{t-1}$ that are marginalized. A subset $x_t$ specifies information about $Z_{t-1}$ and $Z_{t+1}$ to the extent that conditioning on $x_t$ constrains the state of $Z_{t-1}$ and $Z_{t+1}$ compared to its unconstrained probability $\pi(Z_{t\pm1})$ (see (Oizumi et al., 2014; Tononi, 2015; Albantakis and Tononi, 2019) for details). By constraining a subset $Z_{t-1}$, $x_t$ specifies information about its possible cause within the system. Likewise, by constraining a subset $Z_{t+1}$, $x_t$ specifies information about its possible effect within the system[1].

***Integration:*** All subsets $x_t \subseteq s_t$ may specify their own information about other subsets $Z_{t\pm 1}$ (see composition). However, a subset only contributes to the intrinsic information of the system if this information is irreducible ($\varphi(x_t) > 0$). This is tested by partitioning the cause/effect repertoire $\pi(Z_{t\pm 1}|x_t)$ into two parts $\pi(Z_{1,t\pm 1}|x_{1,t}) \times \pi(Z_{2,t\pm 1}|x_{2,t})$ and measuring the difference between the intact and partitioned distributions (see (Oizumi et al., 2014) for details). Of all such partitions $\psi$, the one that makes the least difference to the cause/effect repertoire (termed "MIP" for minimum information partition) determines the integrated information $\varphi(x_t, Z_{t\pm 1})$ specified by $x_t$ over the subset $Z_{t\pm 1}$. Moreover, to be a mechanisms within the system the subset $x_t$ must specify information about its causes and effects, requiring that $\min_{t\pm 1}\left(\varphi(x_t, Z_{t\pm 1})\right) > 0$.

Within system $ABC$ in state (1,0,1), the information that subset $AB_t = (1,0)$ specifies about its causes is reducible, as its cause repertoire $\pi(AC_{t-1}|AB_t = (1,0))$ can be partitioned into $\pi(C_{t-1}|A_t = 1) \times \pi(A_{t-1}|B_t = 0)$. Likewise, the information that $ABC_t = (1,0,1)$ specifies about its effects is reducible, as its effect repertoire $\pi(ABC_{t+1}|ABC_t = (1,0,1))$ can be partitioned into $\pi(C_{t-1}|A_t = 1) \times \pi(AC_{t-1}|BC_t = (0,1))$.

***Exclusion:*** Finally, $x_t$ may specify integrated information $\varphi(x_t, Z_{t\pm 1})$ about various subsets $Z_{t\pm 1}$ within a system. The causal role it plays within the system is determined by the subsets $Z^*_{t-1}$ and $Z^*_{t+1}$ over which $x_t$ specifies the maximal amount of integrated information. $Z^*_{t-1}$ and $Z^*_{t+1}$ are respectively termed the *cause* and *effect purview* of $x_t$. In summary, the amount of integrated information $\varphi(x_t)$ specified by the subset $x_t$ can be expressed as:

$$\varphi(x_t) = \min_{t\pm 1}\left(\max_Z\left(\min_\psi\left(D\left(\frac{\pi(Z_{t\pm 1}|x_t)}{\psi(\pi(Z_{t\pm 1}|x_t))}\right)\right)\right)\right). \qquad (4)$$

The set of all irreducible mechanisms within the system, their cause and effect purviews, and their integrated information $\varphi(x_t)$ compose the intrinsic cause-effect structure of a system in a state $C(s_t)$[2].

## Comparing the macro and micro cause-effect structures

As shown in Figure 3C, the macro cause-effect structure of $ABC_t = (1,0,1)$ is composed of five mechanisms with $\varphi(x_t) > 0$, all first order mechanisms and two higher order mechanisms. The information specified by these mechanisms corresponds to the compositional intrinsic information that the system $ABC$ in state (1,0,1) specifies about itself from the intrinsic perspective. For example, Element $A_t = 1$ specifies that $B_{t+1} = 1$. Likewise, $B_t = 0$ specifies that $C_{t+1} = 1$, but only with $p = 0.75$. Together, $AB_t = 10$ specify $BC_{t+1} = 11$ with certainty ($p = 1.0$). $AB_t = 10$ thus specifies *irreducible* information about the next state of $BC_{t+1}$ that cannot be accounted for by $A_t = 1$ and $B_t = 0$ taken independently. Nevertheless, some of the information specified by the mechanisms within the system may seem redundant from the extrinsic perspective. In our example, $AC_t = 11$ specifies the next state of the system $ABC_{t+1}$ with certainty. As outside investigators, we can thus infer the state of every subset of $ABC_{t+1}$. Note, however, that such an inference requires a mechanism to be performed. The system itself only has information about subsets of $ABC_{t+1}$ if other mechanisms exist, such as $A_t = 1$ or $BC_t = 01$, that specify that particular information (Albantakis and Tononi, 2019).

---

[1] Within the cause and effect repertoire, we can identify the specific state that is maximally constrained by $x_t$, which then corresponds to the specific cause or effect of $x_t$ within the system from the intrinsic perspective of the system (Haun and Tononi, 2019), (Barbosa et al, forthcoming).

[2] In addition, it is possible to evaluate *relations* between the cause and effect purviews of the various mechanisms, which specify the causes and effects specified by multiple mechanisms (see (Haun and Tononi, 2019)).

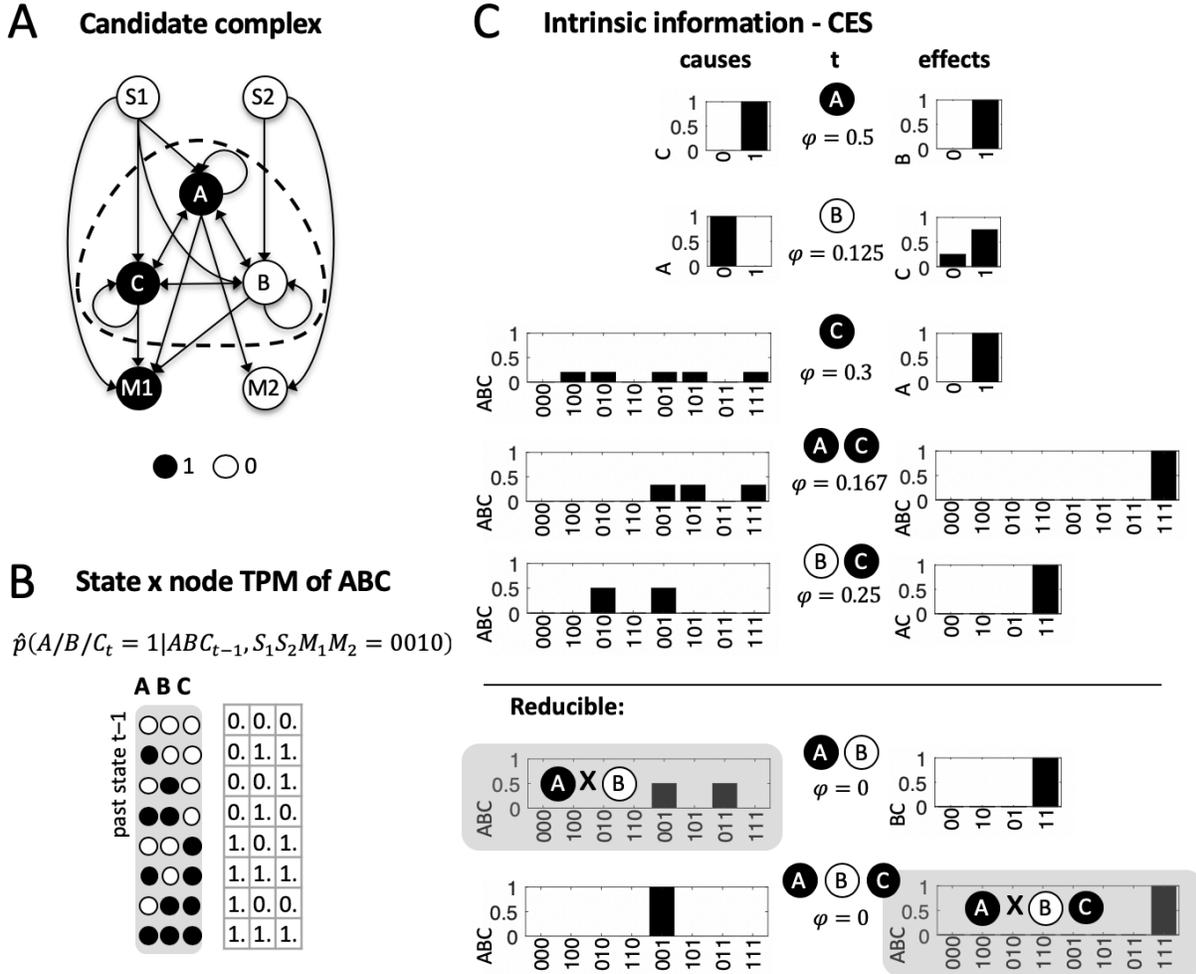

*Figure 3. Compositional cause and effect information of subsystem ABC. (A) The goal is to evaluate the causal information that ABC in its current state specifies about itself, treating the other elements as fixed background conditions. (B) The TPM for subsystem ABC can be obtained from the system's TPM (Figure 1C) by conditioning the full TPM on the current state of $S_1 S_2 M_1 M_2 = (0,0,1,0)$. Since the elements are binary, we can write the TPM in state-by-node format where each column specifies the probability of A, B, or C to be 'on' (1) given the respective input row. (C) Every subset of $ABC_t = (1,0,1)$ may form a separate mechanism in ABC and thus specify information about its possible causes and effects within the system in a compositional manner. Here, the information that subset $AB_t = (1,0)$ specifies about its causes is reducible to a partition of $AB_t$ into $A_t \times B_t$. Likewise, the information that $ABC_t = (1,0,1)$ specifies about its effects is reducible to a partition of $ABC_t$ into $A_t \times BC_t$. The cause-effect structure (CES) of ABC in state $(1,0,1)$ is thus constituted of five irreducible mechanisms and the information they specify.*

The corresponding set of micro elements consists of the 43 elements included in the black boxes $A, B$, and $C$. All other micro elements are taken to be background conditions. The micro system state is the one shown in Figure 2. The micro cause-effect structure is computed based on the micro TPM of the system. All of the 43 micro elements specify first order mechanisms in their current state. In principle, the $\varphi$ values of all subsets of the 43 micro elements would have to be evaluated for higher order mechanisms. However, a set of elements $x_t$ can only form a higher order mechanism if each of the elements shares inputs and outputs with other elements in the set. Otherwise, $\varphi(x_t)$ is necessarily 0 as either the cause or effect repertoire can be partitioned without loss (Oizumi et al., 2014; Mayner et al., 2018). As the connectivity at the micro level is rather sparse, modular, and feedforward, only a few layers of nodes may give rise to higher order mechanisms. We identified 12 higher order mechanisms, one in the input layer of black box $C$, the other 11 are specified by four elements in black box $B$ (second layer, 2-5). On average, the $\varphi$ value of the micro mechanisms is lower than that of the macro mechanisms: $\langle\varphi\rangle_{micro} = 0.16 < \langle\varphi\rangle_{macro} = 0.27$, which means that the macro mechanisms constrain their respective inputs and outputs more than the micro mechanisms.

## Micro and macro system-level integrated information $\Phi$

The cause-effect structure $\mathcal{C}(s_t)$ contains all the intrinsic information the system specifies about itself at the respective level of description. However, the notion of intrinsic information requires that there *is* a system in the first place, meaning one "whole" as opposed to multiple separate sets (Oizumi et al., 2014; Albantakis, 2018; Albantakis and Tononi, 2019). The next step in IIT's causal analysis is thus to evaluate whether and to what extent $\mathcal{C}(s_t)$ is integrated, i.e., irreducible under a partition $\Psi$ of the system. This is quantified by $\Phi(s_t)$, the integrated information of the system as a whole $S$ in a particular state $s_t$:

$$\Phi(s_t) = \min_{\Psi}\left(D\left(\mathcal{C}(s_t); \mathcal{C}(\Psi(s_t))\right)\right). \tag{5}$$

Again, we search for the system partition $\Psi$ that makes the least difference to the cause-effect structure $\mathcal{C}(s_t)$, the MIP (minimum information partition). As defined in (Oizumi et al., 2014; Tononi, 2015), system partitions are unidirectional, rendering the connections from one part of the system $X \subset S$ to the rest ineffective.

If the system does not form a unified whole and can be partitioned into two or more parts without loss, $\Phi = 0$. Also systems in which two or more parts of the system are connected in a feedforward manner cannot be integrated ($\Phi = 0$). In a system with $\Phi > 0$, all parts of the system constrain and are being constrained by the rest of the system above a background of external influences. $\Phi$ can thus be viewed as a measure of how much a system exists for itself, in causal terms.

For $\Phi$ to be high, every possible partition must affect the integrated information $\varphi$ specified by many mechanisms within the system. At the micro level, we identified the MIP as indicated in Figure 4A between the fourth input element of black box $B$ and its one output, the first AND gate in the second layer[3]. As only two first order mechanisms are affected, this cut leads to a comparatively low value of $\Phi(s_t^m) = 0.032$ (the 'm' superscript indicates the micro level, below 'M' stands for macro level).

---

[3] Note that the IIT python package Pyphi cannot, at the moment, compute the $\Phi$ value of a 43 element system exhaustively. To identify the micro MIP and assess $\Phi(s_t^m)$, we took advantage of the modularity of the micro system

While the cause-effect structure at the macro level is based entirely on the macro TPM, the system level integrated information $\Phi(s_t^M)$ is still evaluated by partitioning between micro elements (Marshall et al., 2018). This means that the same set of partitions $\Psi$ is tested and can be compared at the macro and micro level. In this way it becomes impossible to trivially increase the system's integration at certain macro levels by "hiding" weak connections inside the macro elements. The macro TPM of the partitioned system is obtained by black-boxing the partitioned micro system using the same element and state mapping as for the unpartitioned system. Compared to the micro level, the same partition has more substantial effects on the macro cause-effect structure, affecting the effect information specified by $A_t = 1$ and also the cause information specified by $AB_t = 10$ and $BC_t = 01$. For this reason, the integrated information specified by the system at the macro level amounts to the higher value of $\Phi(s_t^M) = 0.213$ (Figure 4B).

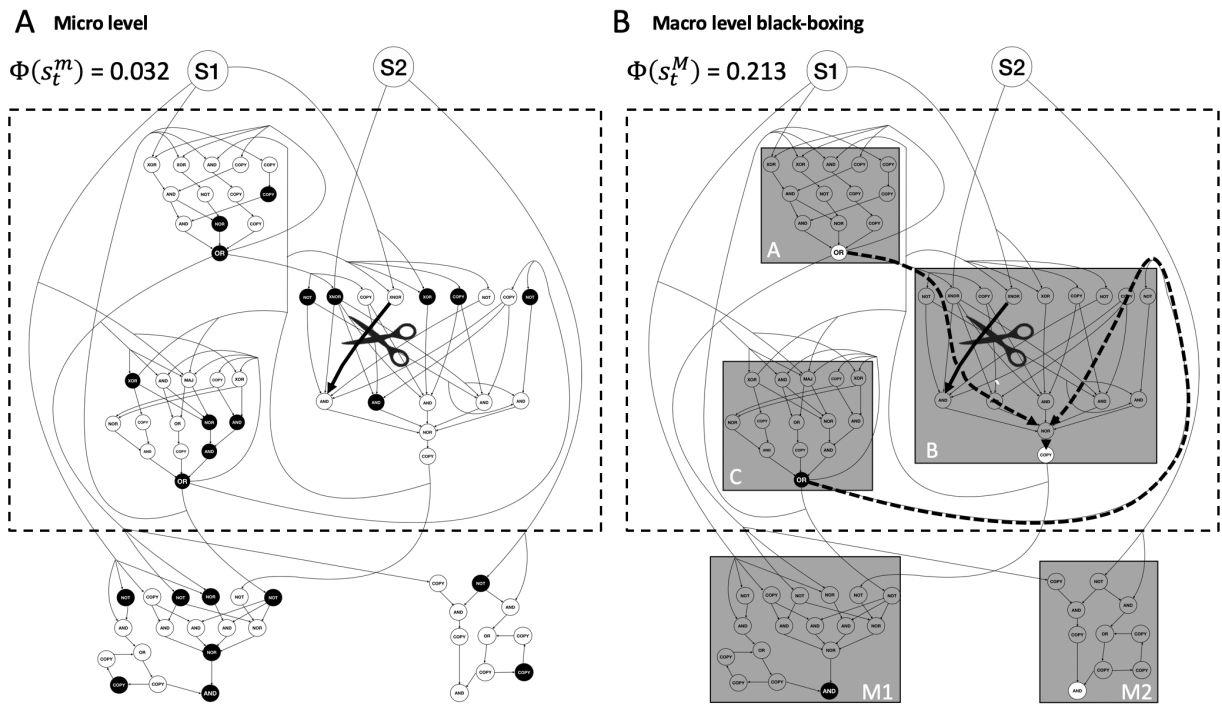

***Figure 4. Comparing the integrated information of the micro and macro level.*** *The same minimal partition (indicated by the scissors cutting the connection in bold) affects the cause-effect information at the micro level less than at the macro level. While in (A) only the two micro elements directly connected by the partitioned arrow are affected, in (B) the partition also has secondary effects on the constraints of macro node A and C on macro node B (as indicated by the dashed, bold arrows). This explains the higher $\Phi$ value at the macro level.*

---

and identified a separate MIP for each black box using Pyphi. The system MIP then corresponds to the minimum across black boxes. Partitions between black boxes all have larger effects on the micro cause-effect structure, as the output nodes of each black box are connected to many micro elements within the system.

According to IIT, maxima of integrated information Φ define causal entities having causal borders with their environment (Oizumi et al., 2014; Marshall et al., 2017, 2018). To identify whether a particular set of elements specifies a maximum of Φ, in principle, requires evaluating many other candidate systems. In our example, all systems larger than the set of elements that constitute A, B, and C (within the dashed rectangle in Figure 4) necessarily have $\Phi = 0$ because the sensors and all elements in $M_1$ and $M_2$ are only connected to ABC in a unidirectional manner). As explained above, systems in which one part is connected to the rest in a feedforward manner cannot form an integrated system according to IIT. Consequently, only sets of elements that are strongly connected (for which a directed path exists from each element to every other element) can have a value of $\Phi > 0$. Using this short-cut, we can establish that the system ABC, as well as the set of its constituting 43 micro elements analyzed in Figure 4A and B form a maximum of integrated information Φ in their current state. This means that removing or adding any element from the set would lead to a lower Φ value[4]. In this way, ABC and its set of constituents define a causal border that separate the internal constraints within the animat from its environment, both at the micro and macro level (Marshall et al., 2017, 2018)[5]. As Φ measures the irreducible intrinsic constraints of a set of elements onto itself over a background of external influences, the macro system ABC can be said to be more autonomous than the set of its micro constituents.

## Tracing back the causal chain leading up to the animat's actions

So far, we have focused on the causal information that the system's elements specify about each other, alone and in combination, at a micro and macro level of description (Figure 3), and its integration as measured by Φ. As we have demonstrated, the macro level description, while supervening on the micro constituents, specifies more integrated information. In particular, the output nodes of the black boxes A, B, and C, play a crucial role in integrating the network over longer time scales.

The causal principles of IIT (such as composition, information, integration, and exclusion), can also be employed to identify and quantify the actual causes and effects of an occurrence ("what caused what"), such as an agent's actions (Albantakis et al., 2019; Juel et al., 2019). An "occurrence" here simply denotes a set of elements in a particular state: $x_t$.

As described above, from the intrinsic perspective, the causal role that an occurrence $x_t$ plays within the system is determined by the causal information $x_t$ specifies about its cause and effect purviews, $Z^*_{t-1}$ and $Z^*_{t+1}$, which are the system subsets over which the amount of integrated information $\varphi(x_t)$ is maximized (Equation 4). The actual state of the cause or effect purviews $Z_{t-1}^*$ and $Z_{t+1}^*$, however, is unknown from the intrinsic perspective of the system in its current state.

---

[4] Our focus here lies on autonomy and the notion of self-defined causal borders that separate the internal mechanisms of the agent from its environment. According to IIT, a physical substrate of consciousness must specify a *global* maximum of Φ across all overlapping sets of elements and spatio-temporal scales. In other words, any particular micro element can only contribute its causal power to one physical substrate of consciousness by IIT's exclusion postulate. Given the size of the animat's micro implementation, an exhaustive analysis across all possible sets of elements and spatio-temporal mappings was not feasible. Thus, it is possible that smaller subsets within the 43 micro elements may specify even higher values of Φ within the system. Likewise, other spatio-temporal mappings may reveal additional "meso" levels of description with higher values of Φ than $ABC_t$.

[5] This also means that the animat's sensors and motors technically form part of the environment, while the causally autonomous entity is defined as the integrated core of the animat.

To identify the *actual* cause $z_{t-1}^*$ of an occurrence $x_t$, instead, we take the perspective of an extrinsic observer of the system with access to the system's time series $\{s_{t-k}, \dots, s_{t-1}, s_t\}$ (Figure 5A). In parallel to $\varphi(x_t)$, the actual cause $z_{t-1}^*$ is then identified as the sub-state $z_{t-1} \subseteq s_{t-1}$ over which $x_t$ specifies the most irreducible causal information:

$$\alpha(x_t) = \max_z \left( \min_\psi \left( \log_2 \left( \frac{\pi(z_{t-1}|x_t)}{\psi(\pi(z_{t-1}|x_t))} \right) \right) \right). \tag{6}$$

$\pi(z_{t-1}|x_t)$ here denotes the probability of the specific state $z_{t-1}$ in the cause repertoire $\pi(Z_{t-1}|x_t)$. The goal is to identify what caused $x_t \subseteq s_t$ given a particular state transition $s_{t-1} \succ s_t$. We refer to the original publication (Albantakis et al., 2019) for further details on the measure $\alpha(x_t)$ and the set of permissible partitions $\{\psi\}$. In deterministic systems, the identified actual cause $z_{t-1}^*$ typically corresponds to an occurrence at $t-1$ that is minimally sufficient for $x_t$ to occur, at least in the case where $x_t$ is a first-order occurrence (a single element in its particular state)[6].

The actions of our example agent are defined by the state of both of its motor units, the output nodes of $M_1$ and $M_2$. As indicated by the micro time series displayed in Figure 5A, the agent's actions are necessarily preceded by a chain of micro events. In the particular state evaluated above $M_1 M_2 = 10$, which means that the animat is moving to the left. In the following we will use "$E_{62}$" and "$E_{72}$" to denote the micro output elements of the black boxes $M_1$ and $M_2$. Both, $E_{62}$ and $E_{72}$ are AND logic-gates and receive direct inputs from two micro elements each, here labeled $E_{59}, E_{61}$ and $E_{67}, E_{69}$ (from left to right). At time $t-1$, these micro elements were in state $E_{59}E_{61}E_{67}E_{69} = 1101$. Applied to the transition $\{(E_{59}E_{61}E_{67}E_{69})_{t-1} = 1101\} \succ \{(E_{62}E_{72})_t = 10\}$, the actual causation analysis here provides the intuitive result that the actual cause of the occurrence $E_{62,t} = 1$ was $(E_{59}E_{61})_{t-1} = 11$ with $\alpha = 2.0$ bits (both inputs had to be 'on' in order to switch the AND-gate $M_1$ 'on') and the actual cause of $E_{72,t} = 0$ was $E_{67,t-1} = 0$ with $\alpha = 0.415$ bits (which prevented $E_{72,t}$ to be 'on'). In principle, we also evaluate if any higher order occurrences specify their own irreducible causes (applying the composition principle). However, in this particular case the occurrence $(E_{62}E_{72})_t = 10$ is reducible, as the elements do not share common inputs at the micro level. $(E_{59}E_{61})_{t-1} = 11$ and $E_{67,t} = 0$ are the direct (or proximal) micro causes of the individual outputs $E_{62,t} = 1$ and $E_{72,t} = 0$. Yet, the animat's action ("move left") here corresponds to the higher-order occurrence $(E_{62}E_{72})_t = 10$, and the proximal micro causes do not provide a causal explanation for why $E_{62,t} = 1$ and $E_{72,t} = 0$ occurred together.

With respect to an agent's action, the direct micro-level cause is rarely considered the cause with the greatest explanatory power (Woodward, 1989). For example, while a motor neuron in the spinal cord may directly initiate a movement, we are typically more interested in identifying the cortical events or external stimuli that triggered the action. To that end, we can employ the actual causation analysis to trace the causal chain of micro occurrences back in time, identifying the "causes of the causes" of the animat's action (Juel et al., 2019). Specifically, we now start with $(E_{59}E_{61}E_{67})_{t-1} = 110$, the union of the actual causes of $(E_{62}E_{72})_t = 10$ and identify the actual causes of all occurrences $x_{t-1} \subseteq (E_{59}E_{61}E_{67})_{t-1} = 110$ at time $t-2$, and so on. As a measure of the causal relevance of a particular micro element, we sum its relative contribution to the $\alpha$ values of all actual causes it participates in within a given time step (see (Juel et al., 2019) for details). Figure 5B shows the results of tracing the causes of $\{(E_{62}E_{72})_t = 10\}$ back to the beginning of the trial ($t = 0$). The histogram on the right shows the summed causal strength across elements. After an initial transient through the micro elements that make up the motor black boxes and the micro elements constituting the internal black boxes A, B, and C ($t = 31$ to $t = 24$, moving upwards from the bottom), a first peak of the overall causal strength can be observed at $t = 23$, when the backtracking reaches the output elements of the black boxes A, B, and C for the second time. This shows that these micro elements play a special causal role, not only with respect to the constraints the system poses onto itself, but also regarding the causes of its actions. Going back further in time, we find additional peaks in correspondence with the spatiotemporal scale that matches the black-box macro level. The reason for these peaks is that the black box outputs act as causal bottlenecks within the system, with many incoming and outgoing connections. Each output element thus contributes to the causes of many occurrences at the next micro time step (setting the states of the many black box input elements).

---

[6] In non-deterministic systems Eqn. (6) would generally identify the minimal occurrence at $t-1$ that raises the probability of $x_t$ the most. Introducing a "specification factor" $\pi(z_{t-1}|x_t)$ in front of $\log_2 \left( \frac{\pi(z_{t-1}|x_t)}{\psi(\pi(z_{t-1}|x_t))} \right)$ in Eqn. 6 effectively implements a tradeoff between an increase in probability of $x_t$ and the cost of setting additional elements into a particular state, which allows identifying the part of $z_{t-1}^*$ that was particularly relevant for the occurrence of $x_t$.

In sum, tracing back the causal chain of events at the micro level of description provides an independent way of identifying nodes within the system that act as "causal bottlenecks". In this way, the actual causation analysis can inform the search for relevant spatiotemporal scales and black boxings that may form maxima of integrated information.

Finally, we can apply the actual causation analysis directly to the macro-level transition $\{(S_1S_2ABC)_{T-1} = 00001\} \succ \{(M_1M_2)_T = 10\}$. In doing so, we find the macro occurrence $(S_1AC)_{T-1} = 001$ (or equivalently $(S_1BC)_{T-1} = 001$) to be the cause of $M_{1,T} = 1$ with $\alpha = 1.30$ bits, and $A_{T-1} = 0$ to be the cause of $M_{2,T} = 0$ with $\alpha = 0.25$ bits[7]. In addition, the higher-order occurrence $(M_1M_2)_T = 10$ specifies its own irreducible cause $(S_1AC)_{T-1} = 001$ at the macro level with $\alpha = 0.35$ bits. This can be interpreted as causal information that the joint occurrence of $(M_1M_2)_T = 10$ specifies about the particular state of $S_1AC$ that actually happened at $T-1$ which is not specified by its parts taken independently. In other words, there is a causal explanation for the action "move left", corresponding to $(M_1M_2)_T = 10$ at the macro level, beyond the independent occurrences of $M_{1,T} = 1$ and $M_{2,T} = 0$. As we have seen above, such an explanation does not exist at the micro level.

---

[7] The $\alpha$ values at the macro level are somewhat smaller than those of the proximal causes at the micro level, as we average over all possible initial states of the motor black-boxes when we evaluate the strength of the causal links in the macro transition, which introduces a certain level of indeterminacy (see (Marshall et al., 2018)).

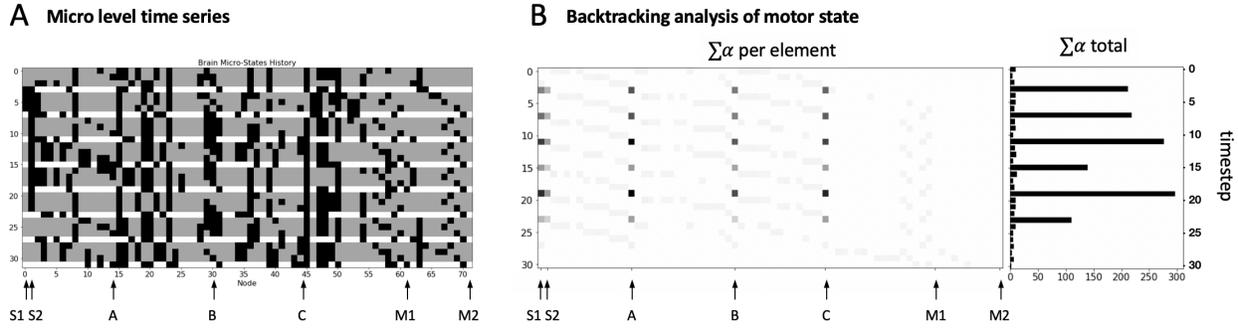

*Figure 5. Tracing back the causes of action $M_1 M_2 = 10$.* (A) Micro level time series specifying the state of all 72 micro constituents of our example animat over 32 time steps. (B) Starting from the current micro state of the agent (here $t = 31$), the actual causes of the occurrences $E_{62} = 1$, $E_{72} = 0$, and $E_{62} E_{72} = 10$ are identified in the preceding micro time-step $t = 30$ (proximal causes). The micro elements $E_{62}$ and $E_{72}$ here correspond to the output elements of the motor black boxes $M_1$ and $M_2$. Iteratively, the backtracking analysis then identifies the causes of the set of elements involved in the proximal causes of the previous time step (causes of causes). At each time step we determine the strength $\alpha$ of the causal link between an occurrence and its actual cause (Eqn. 6) and assign each micro element its summed contribution to $\alpha$ (right panel). The histogram on the left shows the summed contribution across elements. After an initial transient, maxima of causal strength can be observed at every macro time step for the micro elements that correspond to the black box outputs which highlights the special causal role that these output elements play within the system.

## Discussion

In science, macro-level descriptions of the causal interactions within complex, dynamical systems are typically deemed convenient, but ultimately reducible to a complete causal account at the level of the underlying micro constituents. Yet, such a reductionist perspective is hard to square with several properties associated with autonomy and agency that depend on a system's causal structure beyond its individual micro constituents (Ellis, 2016; Albantakis, 2018; Marshall et al., 2018; Albantakis and Tononi, 2019).

For example, the notion of an agent as an autonomous entity that interacts with its environment requires, to begin with, a subdivision of a larger system into agent and environment which cannot be properly formulated using a reductionist account of the system's causal properties (Albantakis and Tononi, 2019). Moreover, while any action performed by an agent is necessarily preceded by a chain of micro events, a dynamical account of "what happened" at the micro level does not equal a causal account of "what caused what" (Albantakis et al., 2019). Nevertheless, the fact that macro-level descriptions of a system supervene upon the dynamics of their micro-level constituents seems to leave no room for genuine macro-level causes and effects (Kim, 1993).

Here we argue that much of the appeal of the reductionist perspective stems from an inadequate notion of causation that is incoherent, fails to account for causal structure, and does not distinguish causation from prediction. By contrast, Integrated information theory (IIT) (Oizumi et al., 2014) offers a consistent, quantitative account of causation based on a set of causal principles, including notions such as causal specificity, composition, and irreducibility, that challenges the reductionist perspective in multiple ways. First, the IIT formalism provides a complete account of a system's causal structure, including irreducible higher-order mechanisms constituted of multiple system elements (Albantakis and Tononi, 2019). IIT's quantitative notion of irreducibility here supplants the reductionist assumption that mechanisms—irreducible causal units—are ultimately restricted to micro elements (Grasso et al., forthcoming). Second, a system's amount of integrated information ($\Phi$) measures the causal constraints a system exerts onto itself and can peak at a macro level of description (Hoel et al., 2016; Marshall et al., 2018). Finally, the causal principles of IIT can also be employed to identify and quantify the actual causes of (higher-order) events ("what caused what"), such as an agent's actions (Albantakis et al., 2019; Juel et al., 2019).

In this chapter, we have demonstrated IIT's causal framework by example of a simulated agent, equipped with a small neural network, that forms a maximum of $\Phi$ at a macro scale. Our particular example agent is constituted of 72 deterministic logic-gates at the micro level, which are grouped into 7 non-overlapping black boxes (including the sensors) at the macro level. The macro-level dynamics supervene upon the micro level and we have full knowledge about the input-output functions of the micro elements and their connectivity. Nevertheless, the agent is an open system, receiving inputs from the environment, which means that its internal dynamics are context dependent and can only be predicted within the larger agent-environment system (Albantakis and Tononi, 2019; Bishop and Ellis, 2020). By construction, in our example the environment updates at the same temporal scale as the black-boxed macro-level. In future work, we plan to investigate which environmental conditions may facilitate the evolution of animats with a hierarchical causal structure, such as environments that update at a slower rate than the animats' micro constituents and other forms or macroscopic constraints (Bishop and Ellis, 2020).

As our example agent is deterministic, the observed increase in intrinsic causal power ($\Phi$) at the macro level is due to the particular causal structure of the system, including higher-order mechanisms and strong constraints across multiple time steps. While the constraints that the system exerts onto itself at the macro level supervene on the micro level, they are intrinsic to the macro level and only become apparent when the system is analyzed as the set of interacting black boxes. In other words, these constraints only exist at the macro level of description. As argued in (Ellis, 2009, 2016; Hoel et al., 2013; Bishop and Ellis, 2020), indeterminism at the micro level may provide additional "causal slack" at the bottom that allows for the macro level to be causally efficacious over the lower levels. In (Hoel et al., 2013, 2016), we have demonstrated that coarse-graining sets of micro elements and their states into macro elements may also increase the causal specificity and integrated information ($\Phi$) at the macro level. In any case, while the macro constraints are not necessary to simulate or predict the future state of the system, they are necessary to explain the stability of the system across multiple time-steps in causal terms.

The same reasoning applies to the agent's actions. The output of the animat's motor units can be simulated, and thus predicted, based on the animat's micro constituents. Nevertheless, the micro level does not offer a causal account of *why* the animat performed this particular action. In our simple example, the micro level, for instance, does not provide a cause for the action as a higher-order occurrence of $(E_{62}E_{72})_t = 10$. By contrast, an irreducible causal explanation exists at the macro spatio-temporal scale.

More generally, the question "*why* an agent chose a particular action over another?" cannot be addressed in purely reductionist terms if the action, the choice, and the agent itself correspond to macroscopic causal structures constituted of many micro occurrences or elements. A principled treatment of notions such as causal autonomy, agency, and free will instead require a quantitative, non-reductionist account of causation.